\input harvmac
\noblackbox

\font\cmss=cmss10 \font\cmsss=cmss10 at 7pt
 \def\inbar{\,\vrule height1.5ex width.4pt depth0pt}
\def\IZ{\relax\ifmmode\mathchoice
{\hbox{\cmss Z\kern-.4em Z}}{\hbox{\cmss Z\kern-.4em Z}}
{\lower.9pt\hbox{\cmsss Z\kern-.4em Z}}
{\lower1.2pt\hbox{\cmsss Z\kern-.4em Z}}\else{\cmss Z\kern-.4em
Z}\fi}
\def\IB{\relax{\rm I\kern-.18em B}}
\def\IC{{\relax\hbox{$\inbar\kern-.3em{\rm C}$}}}
\def\ID{\relax{\rm I\kern-.18em D}}
\def\IE{\relax{\rm I\kern-.18em E}}
\def\IF{\relax{\rm I\kern-.18em F}}
\def\IG{\relax\hbox{$\inbar\kern-.3em{\rm G}$}}
\def\IGa{\relax\hbox{${\rm I}\kern-.18em\Gamma$}}
\def\IH{\relax{\rm I\kern-.18em H}}
\def\II{\relax{\rm I\kern-.18em I}}
\def\IK{\relax{\rm I\kern-.18em K}}
\def\IC{\relax{\rm I\kern-.18em C}}
\def\IR{\relax{\rm I\kern-.18em R}}

\lref\juan{J. Maldacena, ``The Large N Limit of Superconformal Field
Theories and Supergravity," hep-th/9711200.}
\lref\ads{S. Gubser, I. Klebanov and A. Polyakov, ``Gauge
Theory Correlators from Noncritical String Theory,'' 
Phys. Lett. {\bf B428} (1998) 105, hep-th/9802109\semi 
E. Witten, ``Holography and Anti de Sitter Space,'' hep-th/9802150.} 
\lref\ks{S. Kachru and E. Silverstein, ``4d Conformal Field Theories
and Strings on Orbifolds,'' Phys. Rev. Lett. {\bf 80} (1998) 4855, 
hep-th/9802183.}
\lref\kks{S. Kachru, J. Kumar and E. Silverstein, ``Vacuum Energy
Cancellation in a Nonsupersymmetric String,'' hepth/9807076.}
\lref\jeff{J. Harvey, ``String Duality and Nonsupersymmetric Strings,"
hep-th/9807213.}
\lref\senvafa{A. Sen and C. Vafa, ``Dual Pairs of Type II String
Compactification," Nucl. Phys. {\bf B455} (1995) 165, hep-th/9508064.}
\lref\vafawit{C. Vafa and E. Witten, ``Dual String Pairs with
$N=1$ and $N=2$ Supersymmetry in Four Dimensions,'' Nucl. Phys.
Proc. Suppl. {\bf 46} (1996) 225, hep-th/9507050.}
\lref\fhsv{S. Ferrara, J. Harvey, A. Strominger and C. Vafa,
``Second Quantized Mirror Symmetry,'' Phys. Lett. {\bf B361}
(1995) 59, hep-th/9505162.}
\lref\harvstrom{J. Harvey and A. Strominger, ``The Heterotic
String is a Soliton,'' Nucl. Phys. {\bf B449} (1995) 535,
hep-th/9504047.}
\lref\others{S. Kachru and E. Silverstein, ``$N=1$ Dual String Pairs and
Gaugino Condensation," Nucl. Phys. {\bf B463} (1996) 369, hep-th/9511228\semi
O. Bergman and M.R. Gaberdiel, ``A Non-supersymmetric Open String 
Theory and S Duality,'' Nucl. Phys. {\bf B499} (1997) 183, hep-th/9701137\semi
J. Blum and K. Dienes, ``Strong/Weak Coupling Duality and Relations
for Nonsupersymmetric String Theories,'' Nucl. Phys. {\bf B516} (1998) 83,
hep-th/9707160, and ``Duality Without Supersymmetry: The Case of the
$SO(16) \times SO(16)$ String," Phys. Lett. {\bf B414} (1997) 260,
hep-th/9707148\semi
M. Berkooz and S. Rey, ``Nonsupersymmetric Stable Vacua of M-theory,''
hep-th/9807200.}

\Title{\vbox{\baselineskip12pt\hbox{hep-th/9808056}
\hbox{LBNL-42139, SLAC-PUB-7907, UCB-PTH-98/40}
}}
{\vbox{\centerline{
Self-Dual Nonsupersymmetric}\smallskip
\centerline{Type II String Compactifications}}
}
\centerline{Shamit Kachru$^{1}$ and Eva Silverstein$^{2}$ 
}
\bigskip
\bigskip
\centerline{$^{1}$Department of Physics}
\centerline{University of California at Berkeley}
\centerline{Berkeley, CA 94720}
\smallskip
\centerline{and}
\smallskip
\centerline{Ernest Orlando Lawrence Berkeley National Laboratory}
\centerline{Mail Stop 50A-5101, Berkeley, CA 94720}
\medskip
\medskip
\centerline{$^{2}$ Stanford Linear Accelerator Center}
\centerline{Stanford University}
\centerline{Stanford, CA 94309}
\bigskip
\medskip
\noindent
It has recently been proposed that certain nonsupersymmetric
type II 
orbifolds 
have vanishing perturbative contributions to
the cosmological constant.
We show that techniques of Sen and Vafa
allow one to construct dual type II
descriptions of these models (some of which have no weakly coupled
heterotic
dual).  The dual type II models are given by the same
orbifolds with the string coupling $S$ and a $T^2$ volume
$T$ exchanged.  
This allows us to argue that in various 
strongly coupled limits of the original type II models, 
there are weakly coupled duals which exhibit
the same perturbative cancellations as the original models.

\Date{August 1998}

\newsec{Introduction}

Motivated in part by the AdS/CFT correspondence \refs{\juan,\ads}, 
it was recently suggested that certain special nonsupersymmetric
type II string compactifications might nevertheless have 
vanishing cosmological constant $\Lambda$ \ks.  
A concrete candidate example (with supporting perturbative computations at
the one and two loop level, as well as a more heuristic argument
for higher-loop cancellations using the general form of the 
higher genus twist structures) 
was presented in \kks.\foot{Subtleties in the gauge
choice at two loops in \kks\ are being investigated.
At higher loops there is no known consistent method for
fixing the gravitino gauge slice in superstring perturbation theory,
so we are left so far with a more heuristic understanding of
these contributions and how they appear to cancel.}

Unfortunately, explicit checks at higher loops are generally difficult
to perform, and provide no insight into possible nonperturbative corrections
to the vacuum energy.  In a recent paper \jeff, Harvey pointed out that 
techniques from string duality can be fruitfully applied to 
models like the one studied in \kks\ (for some discussions of 
duality and nonsupersymmetric string
vacua in other contexts see e.g. \others).  By computing
at one loop in a dual heterotic model, he finds that at a given
order in a type II $\alpha^\prime$ expansion,
perturbative contributions in the type II coupling 
cancel to all orders, providing a test
of the all-orders vanishing conjectured in \kks.  Moreover,
he finds nonzero non-perturbative contributions.     
The precise model studied in \jeff\ is a slight variant of the model
proposed in \kks; the difference between the models is crucial in
enabling one to construct a heterotic dual for the modified model. 

In this note, we point out that both models of the type studied in
\kks\ and the modified version discussed in \jeff\ 
(as well as some new models which share their interesting features)
can also be
given dual type II descriptions in certain limits.  
In fact, these models are very 
special -- they turn out to be $\it self-dual$ as type II vacua. 
By this we mean that the 
models with given values of the coupling
$S$ and a $T^2$ volume modulus $T$ are dual to the 
same orbifolds, with $S$ and $T$ interchanged. 
This allows us to argue that in various strongly coupled regimes of 
the original theory, there exists a dual weakly coupled 
type II theory which also has vanishing leading perturbative contributions
to the cosmological constant.
Moreover, in the specific model of \kks\ as well as in the new
models to be introduced here, there is no weakly coupled heterotic limit and
all type II duals enjoy the same perturbative cancellations that 
occur in the original model.
This does not imply that $\Lambda$ vanishes exactly in the models without
heterotic duals,  
but it does mean that any concrete checks we can perform using
string duality (along the
lines of \jeff) are consistent with such vanishing.

In \S2, we briefly review the construction \senvafa\ which allows us 
to systematically determine the type II duals.  
In \S3, we apply this construction to the models of interest and find that
they are self-dual.  The implications of this finding for
the models which do and do not admit heterotic duals are described.
We close with a brief discussion in \S4.

\newsec{Brief Review of Type II Orbifold Dual Pairs}

\subsec{Basic Idea}

A systematic construction of dual pairs of type II compactifications in 
$D=4,5$ dimensions was discussed by Sen and Vafa \senvafa.  
The U-duality group of type II compactifications on $T^4$ is
$SO(5,5;\IZ)$, while the perturbatively obvious (T-duality) subgroup
is $SO(4,4;\IZ)$.  
Consider two elements $h,\tilde h \in SO(4,4;\IZ)$ of order $n$,
which are not conjugate in $SO(4,4;\IZ)$ but which $\it are$ conjugate
in $SO(5,5;\IZ)$:

\eqn\conjugate{g h g^{-1} = \tilde h,~~g \in SO(5,5;\IZ)}

\noindent
There is a subspace $\cal{M}$ 
of the moduli space of $T^4$ compactifications
which consists of theories with 
an extra $h$ symmetry (which one finds by looking for
fixed points of the $h$ action on the Teichm\"uller space).  
Using the U-duality transformation $g$, this is dual to
the subspace $\tilde{{\cal{M}}}$ invariant under $\tilde h$.
Since $g$ is not in $SO(4,4;\IZ)$, the duality is not obvious
perturbatively.
On ${\cal M}$ and $\tilde{{\cal M}}$, $h$ 
and $\tilde h$ are realized as symmetries of
the compactifications in question.
  
Now, consider
compactifying on an additional $T^2$, which we can
take to be a product of two circles.  We 
can orbifold the resulting 
compactifications by $h$ ($\tilde h$) acting on the
$T^4$ combined with a free 
order $n$ action on the $T^2$. 
By the adiabatic argument \vafawit, the resulting models in
four dimensions are still dual.

\subsec{Element of Interest}

The particular $SO(5,5;\IZ)$ element $\bar g$ of interest is
given in 
terms of the element $\sigma$ of the ten-dimensional $SL(2,\IZ)$ 
symmetry group of type IIB which inverts the ten-dimensional
axion/dilaton field and the T-duality element $\tau_{1234}$ that inverts
the volume of the $T^4$:  
\eqn\barg{{\bar g} ~=~\sigma \cdot \tau_{1234}\cdot \sigma^{-1}}
This maps fundamental strings (without winding on the $T^4$) to
NS fivebranes wrapped on the $T^4$.  
The element $\bar g$ has the very helpful property that
\eqn\helpful{{\bar g} h {\bar g}^{-1}  \in SO(4,4;\IZ)}
for all $h \in SO(4,4;\IZ)$ \senvafa.

One interesting and important property of 
the element $\bar g$ is that, as long as the Ramond-Ramond scalar
VEVs in $D=6$ vanish, it acts on the dilaton and the three-form field
$H_{\mu\nu\rho}$ in exactly the same way as the
type IIA / heterotic string-string duality.
Therefore, after the further compactification on an (orbifolded)
$T^2$, the dual models that the adiabatic argument yields will be
related by $S-T$ exchange:

\eqn\duality{\tilde S ~=~ T,~~\tilde T ~=~ S}

\noindent
where $T$ is the K\"ahler modulus associated with the $T^2$ and $S$ is the
axion-dilaton in four dimensions.  This map is inherited from the
supersymmetric theory where certain quantities are holomorphic in 
$S$ and $T$.  In the nonsupersymmetric theory nothing is determined
by holomorphic objects, and generic computations will depend on both
$S$ and $S^\dagger$ as well as $T$ and $T^\dagger$. 

By detailed considerations we will not repeat here, Sen and Vafa
derive the result of conjugating various elements of
$SO(4,4;\IZ)$ by $\bar g$. 
Consider an element 
$h \in SO(4,4;\IZ)$ that acts on the 8-vector $X^{1...4}_{L},
X^{1...4}_{R}$
as
\eqn\hact{h ~=~\pmatrix{\omega(\theta_L) & ~ & ~ & ~\cr
                       ~ & \omega(\phi_L) & ~ & ~\cr
                       ~ & ~ & \omega(\theta_R) & ~\cr
                        & ~ & ~ & \omega(\phi_R)}}
where
\eqn\rot{\omega(\theta) ~=~\pmatrix{cos(\theta) & sin(\theta)\cr
                                       -sin(\theta) & cos(\theta)}}
We will abuse notation and denote elements of $SO(4,4;\IZ)$ which
are like \hact\ by $(\theta_L,\phi_L,\theta_R,\phi_R)$. 
Then ${\bar g}$ conjugates $h$ to $\tilde h$ which acts on
$X^{1...4}_L, X^{1...4}_R$ as $(\tilde \theta_L, \tilde \phi_L, 
\tilde \theta_R, \tilde \phi_R)$
where

\eqn\dual{\pmatrix{\tilde \theta_L \cr
                   \tilde \phi_L \cr
                   \tilde \theta_R \cr
                   \tilde \phi_R}
~=~ \pmatrix{1/2 & -1/2 & 1/2 & -1/2 \cr
             -1/2 & 1/2 & 1/2 & -1/2 \cr
             1/2 & 1/2 & 1/2 & 1/2 \cr
             -1/2 & -1/2 & 1/2 & 1/2}
~\pmatrix{\theta_L \cr
          \phi_L \cr
          \theta_R \cr
          \phi_R}}

\noindent
\dual\ is the equation that will yield the type II duals of our orbifolds.
We will also discuss directly  
how the map arises in our particular examples.

\newsec{Application to Nonsupersymmetric Models}

\subsec{Original Model}

The model of \kks\ is constructed by orbifolding type II on a $T^6$
consisting of a product of six circles.  The orbifold group
has two generators
$f$ and $g$: 
\bigskip
\vbox{\settabs 3 \columns
\+$S^1$&$f$&$g$\cr
\+1&$(-1,s)$&$(s,-1)$\cr
\+2&$(-1,s)$&$(s,-1)$\cr
\+3&$(-1,s)$&$(s,-1)$\cr
\+4&$(-1,s)$&$(s,-1)$\cr
\+5&$(s^2,0)$&$(s,s)$\cr
\+6&$(s,s)$&$(0,s^2)$\cr
\+&$(-1)^{F_R}$&$(-1)^{F_L}$\cr} 

\noindent
Above, we have indicated how each element acts on the left and right
moving RNS degrees of freedom of the superstring.  $s$ refers to a shift
by $R/2 = l_s / 2\sqrt{2}.$

If we concentrate on the action on the first four circles, then we see
that in the notation of \S2.2\ $f$ and $g$ can be represented as

\eqn\fandg{f ~=~ (\pi,\pi,2\pi,0),~~g ~=~ (2\pi,0,\pi,-\pi)} 

\noindent
(where we use the fact that e.g. $(-1)^{F_R}$ can be represented by a 
$2\pi$ rotation on right movers).  
Then from the action of \dual,
we see that $\tilde f = f$ and $\tilde g = g$.  
So after composing with the further action on the $X^{5,6}$ $T^2$, 
we will find that
the $f$ orbifold is self-dual as is the $g$ orbifold, and so is 
the orbifold by both $f$ and $g$. 

This self-duality can be understood in terms of the duality between
the fundamental string and the wrapped NS fivebrane mentioned above.
As explained in \harvstrom, there are normalizable RR 
zero modes of the NS fivebrane associated to the 2-forms 
on the $T^4$.  These zero modes correspond to the worldsheet
embedding coordinates in the dual type II string.  The
element $f$ in the original model kills half the RR fields.
This means that the dual element $\tilde f$ must act with
a $(-1)$ on half the 8 worldsheet scalars $\tilde X^i_L, \tilde X^i_R,
i=1,\dots,4$ 
associated with
the dual $T^4$. It also   
preserves 1/4 of the supersymmetry.  Since it preserves
some supersymmetry, the dual element $\tilde f$ must act
with a $(-1)$ on only the left-movers or only the right-movers.
To kill 3/4 of the supersymmetry it must also involve an
action $(-1)^{F_R}$ or $(-1)^{F_L}$ respectively.  Thus
$\tilde f$ is isomorphic to $f$.  
The element $\tilde g$ has similar properties, but kills
the rest of the supersymmetry, so it must be isomorphic to $g$.  

We have not yet discussed the duals of the shifts involved in
our group elements.  As in \jeff, we must choose them to
level-match.  These shifts map to gauge transformations
of the RR fields in the NS fivebrane background (as was
first discussed in \fhsv).  Such gauge transformations 
only have a non-perturbative
effect on the dual side.

Now \duality\ tells us that we can do some strong coupling
computations in the original theory by doing small radius 
(of the $T^2$) computations
in the dual.  On the other hand, the dual coupling is related to
the K\"ahler modulus $T$ of the $T^2$ in the original theory.  Therefore, 
one can analyze the theory in several different regimes: 

\noindent
1) If the original theory is weakly coupled ($S \rightarrow
\infty$) and at arbitrary $T$, then the analysis of \kks\ goes
through and at least the leading perturbative contributions to
$\Lambda$ as a function of $S$ vanish.  

\noindent
2) Next let us look at the dual model in its perturbative regime,
i.e. $\tilde S\to\infty$ and arbitrary $\tilde T$.  As just discussed,
this model is isomorphic to the original model.  Therefore in
this regime, perturbatively (at least to two loops) 
in $\tilde S=T$ and to all orders and non-perturbatively
in $\tilde T=S$ (at these orders in $\tilde S$), we find no contribution.   
This is in contrast to the situation in \jeff, where a heterotic
dual was obtained in this limit which also had no contributions perturbatively
in $\tilde T$ but did have nonzero contributions non-perturbatively.  
In both cases one finds a test of the proposal that all perturbative
contributions in $\tilde T=S$ should cancel.

\noindent
3) Finally, if the original model is strongly coupled and
at small radius ($S, T \rightarrow 0$), then we should first
T-dualize to get to strong coupling at large radius $T \rightarrow \infty$.
Then, we can use the duality \duality\ to get the dual model at
weak coupling $\tilde S = \infty$ and small radius. 
Then, since the model is self-dual and we are at weak coupling, 
we are back in situation 1) and the
analysis of \kks\ applies. 

The upshot is that \duality\ implies that in 
all these limits in $S-T$ space, there is a weakly coupled dual.  
Because the model is actually self-dual, 
the perturbative evaluation 
of $\Lambda$ in the dual vanishes at the leading orders of perturbation
theory exactly as in \kks. 

\subsec{Some New Models}

In \kks\ we emphasized the non-abelian nature of the orbifold as
a simple way to ensure that the 1-loop vacuum energy would cancel.
In fact, as noted also in \jeff, this was not actually necessary
in these models.  Nonzero contributions could only occur in contributions
to the 1-loop (torus) amplitude which involve twisting by 
group elements on the $(a,b)$ cycles of the torus which break all
of the supersymmetry. 
With the non-abelian structure, these never occur.  If we remove
the asymmetric shifts along the $T^4$ (and adjust the asymmetric
shift along the 5th circle so as to satisfy level-matching), 
then $f$ and $g$ would
commute, so the $(f,g)$ twist would need to be taken into account
(along with various other supersymmetry breaking twists that can
be obtained from $(f,g)$ by modular transformations, like
$(f,fg)$ and $(g,fg)$). 
This twist structure describes a trace over $f$-twisted states 
with an insertion of the $g$ operator.  It is simplest to
consider the spectrum in light-cone gauge.  In the $f$-twisted
sector, the vacuum after GSO projection can easily be seen to
have equal numbers of $g$-invariant and $g$-anti-invariant states.
The right-moving Ramond vacuum is a spinor, and
the right-moving NS vacuum is of the form 
$\psi^{1,\dots,8}_{-1/2}\vert -1/2 \rangle$.  The four reflections in
$g$ kill half of each of these sets of states.  
This ensures that all mass levels have this property, and the
contribution cancels, though the diagram has information about
the full supersymmetry-breaking in the model.  Similar remarks
apply of course to the other
supersymmetry breaking twists at 1-loop, which can be obtained
from this by modular transformations.

This said, we now can consider another model where
we remove the reflections in
$g$ and change the shifts to
a single symmetric shift on $X^6$.  Then again in 
the $g$-twisted sector, the $f$ operator
has equal numbers of $f$-invariant and $f$-anti-invariant states.
So the $(g,f)$ (and other modular equivalent) supersymmetry breaking 
twist structure contributions
vanish.  One can also use U-duality to find type II duals which
give one control over various strong-coupling limits of these
models,  
as in \S3.1.

\subsec{Models with Heterotic Duals}

Finally we can study various limits of the model discussed in \jeff.
This model has five noncompact dimensions, arising from a compactification
on the orbifold $(T^4\times S^1)/\{f^\prime,g^\prime\}$.  
The element $f^\prime$ differs from $f$ above in that there
is no shift on the sixth circle.  The element
$g^\prime$ differs from $g$ in that its asymmetric
shift $(0,s^2)$ acts on the fifth circle and there is
no extra symmetric shift.  This means that the element
$f^\prime g^\prime$ (combined with a full lattice shift)
creates a K3.   
The $T^4$ is square and
at the self-dual radii while the $S^1$ has a variable radius $R$.
We will also keep track of the six-dimensional string coupling $\lambda$.   
If we start with the type IIA string, 
in the limit ($\lambda_A \to \infty, R_A \to \infty$) there
is a weakly-coupled heterotic dual description \jeff.  Let us
consider now the limit $\lambda_A \to \infty, R_A \to 0$.  
We can T-dualize this
to type IIB in the limit $\lambda_B\to\infty, R_B\to\infty$ where
the subscript $B$ denotes the corresponding quantities
$R_B=\alpha^\prime/R_A$ and $\lambda_B=\lambda_A\sqrt{\alpha^\prime}/R_A$
in the IIB theory.  The IIB theory on K3 has the 10d S-duality
symmetry $\sigma$ as well as the volume inversion symmetry
$\tau_{1234}$ that enter in $\bar g$ \barg.  This means that
we can use this element to generate a type II/type II string-string dual in
this regime for this model.  Perturbative contributions to the
cosmological constant in this regime will cancel as in the original
model.  

If duality is a valid technique for studying these
models, the heterotic limit \jeff\ implies that there is a nonvanishing
dilaton potential (nonperturbative in the original type II coupling). 
As one shrinks $R_A$ (moving from the regime of validity of the
heterotic dual towards the regime of validity of the type II dual),
the heterotic dual develops a tachyon.  It is therefore not 
completely clear that
the heterotic and type II duals are connected by changing
parameters in the nonperturbative string model.  
In any case, the results of \jeff\ seem to imply that in 
cases with heterotic duals the type II duals 
must also have a nontrivial dilaton potential of a form which is
not ruled out by existing calculations.\foot{Very naive extrapolation of
the results of \jeff\ indicate that the dilaton potential would behave
like $e^{-1/\tilde \lambda_B}$ in the dual type IIB theory, because
it behaves
like $e^{-1/R_h}$ at small heterotic radius $R_h$.}  
It is clear however that the models of \S3.1\ and \S3.2\ do not have
a non-perturbative potential at the same order in the
original type II $\alpha^\prime$
and $g_{st}$ as the contribution detected in \jeff.

\newsec{Discussion} 

As we have seen, varying the way the shifts (and to some extent
the reflections) occur in these orbifold models changes some
of their non-perturbative properties significantly.  In particular,
some models have heterotic duals, while others only have type II
duals.  The trick used in \kks\ for cancelling perturbative contributions
to the cosmological constant was specific to type II in that
the freedom to use both $(-1)^{F_L}$ and $(-1)^{F_R}$ was crucial.  
It is therefore reassuring that some models only have type II duals
from the point of view of aiming for non-perturbative cancellation
(although our results do not imply
that there is exact cancellation in
any of the models discussed here, since duality considerations only
allowed us to check for a small subset of the possible nonperturbative
contributions).
An exactly flat dilaton potential in similar 3d models
could be of interest for potentially generating a four-dimensional model
with no cosmological constant upon taking the strong-coupling limit. 
On the other hand, if one considers this type of model in 4d,
one might ultimately want models with non-perturbative
dilaton potentials \jeff, either to stabilize the dilaton or to agree with
observations suggesting a small nonzero vacuum energy.  
Of course we are still far from a truly realistic
model in which to usefully discuss these issues.   

It is interesting that these slight variations in the shifts
used change the dual descriptions so drastically.  As mentioned
in \jeff, it would be very interesting to study the D-brane
spectra in these models in order to understand their degeneracy
(or lack thereof) in the various orbifold formulations.

\centerline{\bf{Acknowledgements}}

We would like to thank J. Harvey, J. Kumar
and G. Moore for interesting
discussions and J. Harvey for helpful comments on a draft
of this note. 
S.K. is supported by
NSF grant PHY-94-07194, by DOE contract DE-AC03-76SF0098, by
an A.P. Sloan Foundation Fellowship and by a DOE OJI Award.
E.S. is supported by DOE contract DE-AC03-76SF00515.

\listrefs
\end